\documentclass[twoside,slac_one]{revtex4}
\usepackage{graphicx}
\usepackage{fancyhdr}
\usepackage{amsmath} 
\usepackage{bm}
\usepackage{amsxtra}
\usepackage{amssymb}
\usepackage{amsthm}
\usepackage{latexsym}
\usepackage{lscape}

\begin{document}

\title{Overview of Lorentz Violation in Neutrinos}

%

\author{Jorge S. D\'\i az}
\affiliation{Physics Department, Indiana University, Bloomington, IN 47405, USA}

\begin{abstract}
A general introduction to Lorentz and CPT violation is presented. The observable effects of the breakdown of such fundamental symmetries in neutrinos are outlined and different experimental searches of some of the key signals of Lorentz violation are discussed. A novel powerful method to study model-independent neutrino oscillations is reviewed and then used to illustrate a new global model of neutrino oscillations based on Lorentz violation consistent with atmospheric, accelerator, reactor, and solar neutrino data. 
\end{abstract}

\maketitle

\thispagestyle{fancy}


\section{Introduction}
The discovery of neutrino oscillations constitutes the first crack in the Standard Model (SM) that not only triggered a new generation of experimental developments but also challenged our theoretical paradigm to describe these fundamental particles. The popular description of neutrino oscillations extends the SM by incorporating tiny masses to the three neutrino flavors using a non-diagonal mass matrix that mixes the three flavors \cite{pdg}. Consequently, neutrino flavors are linear superpositions of states of definite mass that propagate with different frequencies determined by their masses. This standard model of three massive neutrinos (3$\nu$SM) has served as a successful description of established experimental results during the last decades. In more recent years, however, new experimental results seem to challenge this model suggesting the possibility of physics beyond the 3$\nu$SM. Even though these anomalous results remain to be confirmed, the interferometric nature of neutrino oscillations makes them sensitive probes of scenarios in which new physics beyond the SM can be observed.\footnote{the same arguments apply to the oscillations of neutral mesons \cite{NeutralMesons}.} One such scenario is the breakdown of Lorentz invariance \cite{KS89}. 

These proceedings present an overview of the theory of neutrino behavior in the presence of Lorentz and CPT violation as well as discussions about experimental searches based in this theory. General aspects of Lorentz and CPT violation and the framework to study its observable effects are reviewed. The use of a model-independent plot designed to visualize neutrino oscillation signals in different experiments simultaneously is motivated and some key signals of Lorentz and CPT violation in experiments are discussed. The final section, which is based in Refs. \cite{DK1,DK2}, presents a global model for neutrino oscillations based in Lorentz violation as an alternative to the 3$\nu$SM. The model is consistent with established data from accelerator, atmospheric, reactor, and solar experiments using only three parameters.

\section{Lorentz violation}

\subsection{Basics}

During the last two decades, the possibility that Lorentz invariance could not be an exact symmetry of nature has been growing. In 1989, a seminal work by Kosteleck\'y and Samuel indicated that mechanisms in string field theory could produce the spontaneous breaking of Lorentz symmetry \cite{KS89}. 
To understand the concept of Lorentz violation we can use a well known example: in the electroweak theory, a scalar field acquires a nonzero vacuum expectation value (vev) leading to mass terms of fields in the theory coupled to this vev. This breaking of the electroweak symmetry gives mass to many of the particles in the theory. Similarly, in string field theory a tensor field that acquires a vev leads to terms of standard fields in the theory coupled to the tensor vevs. Since these tensor vevs are not scalars (they carry spacetime indices), they correspond to functions in a given frame. This causes the interaction to change depending on the direction or velocity of the fields coupled to these vevs. In other words, these tensor vevs act like fixed background fields causing the breakdown of Lorentz symmetry.

It is important to emphasize that the theory remains independent of the observer frame, this means that even after Lorentz symmetry is broken the theory is invariant under coordinate changes. When a measurement is performed, a comparison between two systems is made. The result of this measurement is independent of the direction or velocity of the observer because when the coordinates of the observer are rotated or boosted then the two systems in comparison transform. This invariance under {\it observer Lorentz transformations} is not related to the concept of Lorentz violation. On the contrary, when properties of the field (such as spin or direction of propagation) are rotated or boosted instead of the observer's coordinates we talk about {\it particle Lorentz transformations}. In vacuum, particle and observer transformations are inversely related; many authors call these transformations active and passive transformations, respectively. Nevertheless, in the presence of a background field these identifications hold only partially. Under an observer transformation, both properties of the particle and the background field transform to a new set of coordinates, just like a passive transformation; nonetheless, under a particle transformation only the properties of the particle transform while the background field remains unchanged. A background field breaks invariance under particle transformations. Notice that in the presence of a background field a particle transformation cannot be identified with an active transformation because in the later all fields, including background fields, transform.

\subsection{Framework}

The observation that there is no compelling experimental evidence of Lorentz violation to date suggests that Lorentz-violating effects must be small and the natural suppression scale to consider is the Planck scale ($M_P\simeq10^{19}$ GeV). Even though the study of high-energy phenomena is limited by our technological capabilities, the search for low-energy suppressed effects and high-precision measurements are possibilities to which we have access today. Starting in the 1990s, the development of theories describing Lorentz violation considered the study of effective interactions \cite{KP1995}. 

In the absence of compelling signals of Lorentz violation a natural approach is the formulation of a general framework that can account for all possible observable effects that could arise due to the breakdown of Lorentz symmetry. In other words, the fact that we have not seen Lorentz violation it does not mean that the symmetry is exact, it could happen that we have not searched for the appropriate signals. If we do not want to miss the possibility of observing effects of the violation of such a fundamental symmetry we have to look everywhere.
Such general framework, called Standard-Model Extension (SME), was proposed in 1997 by Colladay and Kosteleck\'y \cite{SME} as a realistic Lorentz-violating extension of the SM based on effective field theory that is independent of the underlying theory. The SME extends the SM and General Relativity \cite{SME2004} by incorporating coordinate-independent terms in the action that break Lorentz symmetry. These {\it observer scalars} are constructed by contracting operators of standard fields with controlling coefficients. For instance, the fermion sector of the SME includes Lorentz-violating terms in flat spacetime of the form \cite{KP1995,SME,NRterms1,NRterms2}

\begin{equation}
\label{L_LV}
\mathcal{L}_\text{LV} \supset 
\frac{1}{2}\,i \overline{\psi}\, \gamma^\mu\!\!\stackrel{\leftrightarrow}{D_\mu}\!\!\psi
-(a_L)_\mu \overline{\psi} \,\gamma^\mu \psi
+\frac{1}{2}\,i (c_L)_{\mu\nu} \overline{\psi} \,\gamma^\mu\!\!\stackrel{\leftrightarrow}{D^\nu}\!\!\psi
+\ldots,
\end{equation}
where $(a_L)$ and $(c_L)$ correspond to tensor vevs in the underlying theory. We refer to these quantities as {\it coefficients for Lorentz violation}. The ellipsis denotes high-order derivatives that arise when the complete SME is considered \cite{KP1995,SME,NRterms1}.

In flat spacetime, coefficients for Lorentz violation are spacetime constants; however, in a curved spacetimes these coefficients can vary leading to propagating Nambu-Goldstone modes \cite{NGmodes}. The invariance under observer Lorentz transformations is evident because spacetime indices of the coefficients for Lorentz violation and the free spacetime indices of the corresponding operator are properly contracted. These coefficients for Lorentz violation play the role of fixed background fields arising as tensor vevs that break Lorentz symmetry.

\subsection{Physical effects}

We can understand the observable effects predicted by the SME by using a well known system as an analogy. In the Zeeman effect, for instance, the splitting of some spectral lines occurs because a weak magnetic field acts as a background field that sets a preferred direction in space. Particle rotation invariance is broken, in fact a rotation of the spin of the electron respect to the preferred direction would increase or decrease the splitting of the spectral lines; however, invariance under observer transformations is unaffected because a rotation of the coordinates leaves the properties of the system unchanged causing no physical effect. Back to Lorentz violation, coefficients for Lorentz violation play the role of the magnetic field, acting as background fields that break particle Lorentz invariance. Just like in the Zeeman effect, this symmetry breaking produces observable effects that can be studied. Note, however, that since coefficients for Lorentz violation carry spacetime indices, the effects are not only caused by rotation of properties of the field of interest, the breaking of boost invariance introduces isotropic effects.

One of the interesting questions that arises if Lorentz symmetry is spontaneously broken is the fate of the Nambu-Goldstone modes. Studies have shown that when gravity is considered the Nambu-Goldstone modes associated to spontaneous Lorentz violation could play important roles such as the graviton, the photon, or spin-independent forces \cite{NGmodes}.

\subsection{CPT violation}

In the context of realistic effective field theories, CPT violation is accompanied by Lorentz violation \cite{owg}. Since the SME describes general Lorentz violation, a subset of operators in the SME also break CPT. For instance, the second term in Eq. (\ref{L_LV}) is CPT odd; in other words, the operator contracted with the coefficient $a_\mu$ reverses its sign under a CPT transformation.
In general all CPT-odd terms in the theory can be easily identified because the corresponding operator of standard fields contains an odd number of free spacetime indices. Notice that in the SME particles and antiparticles have the same mass. Since a mass term in the lagrangian is scalar under particle Lorentz transformations (a quadratic form of field operators contains no free spacetime indices), this term remains unchanged under CPT. Moreover, if we consider a fermion field whose mass term is given by $m\overline\psi\psi$; both fermion and antifermion are components of the spinor field $\psi$ whose mass is defined as the parameter $m$ in front of the operator $\overline\psi\psi$; therefore, in this context fermion and antifermion cannot have different masses.

The development of the SME allowed experimental searches of Lorentz and CPT violation in different sectors of the SM, triggering worldwide searches of the experimental signals predicted by the SME. Experimental results of all these searches are summarized in the {\it Data tables for Lorentz and CPT violation} \cite{tables}. The SME has also motivated the study of Lorentz violation at the fundamental level \cite{SME_developments}.

\section{Neutrino oscillations}

In a general model-independent description, the change of one neutrino flavor into another as they propagate is the consequence of a non-diagonal hamiltonian not having a complete degenerate spectrum. If the hamiltonian $h_{ab}$ describing neutrino propagation in flavor basis is put in a diagonal form $h=\mathbf{U}^\dag \mathbf{E} \mathbf{U}$ by a unitary matrix of elements $U_{a'a}$, then the probability of a neutrino flavor $\nu_b$ oscillating into a flavor $\nu_a$ is given by\footnote{for simplicity, here we have assumed a real hamiltonian; otherwise, a second term of the form $\sum_{a'>b'} B_{ab}^{a'b'}(U)\, \sin(\Delta_{a'b'} L)$ must be added.} \cite{pdg,KM1}

\begin{equation}
\label{P_ab}
P_{\nu_b\to\nu_a}(L)=\delta_{ab}-\sum_{a'>b'} A_{ab}^{a'b'}(U)\, \sin^2(\Delta_{a'b'} L/2),
\end{equation}
where $L$ is the separation between the neutrino source and detector, the oscillation amplitude $A_{ab}^{a'b'}(U)$ is a function of the elements $U_{a'a}$ of the mixing matrix, $\Delta_{a'b'}=\lambda_{a'}-\lambda_{b'}$ is defined as the eigenvalue differences, and the indices in the diagonal basis take the values $a',b'=1,2,3$.

The conventional description of neutrino oscillations by the 3$\nu$SM uses a hamiltonian parametrized by two mass-squared differences ($\Delta m^2_\odot,\Delta m^2_\text{atm}$), three mixing angles ($\theta_{12},\theta_{13},\theta_{23}$), and one CP-violating phase $\delta_\text{CP}$. One of the key features of this model is that the hamiltonian is proportional to $1/E$, where $E$ is the neutrino energy. The nonzero eigenvalues of this model are $\Delta m^2_\odot/2E$ and $\Delta m^2_\text{atm}/2E$, which lead to oscillation phases in Eq. (\ref{P_ab}) controlled by the ratio $L/E$. This model is very successful describing neutrino oscillation data at low (few MeV) and high energies (few GeV), regimes that show oscillatory signatures as functions of $L/E$ as reported by KamLAND \cite{KamLAND(L/E)} and Super-Kamiokande \cite{SK(L/E)} experiments. 

The success of the 3$\nu$SM has been challenged in recent years by some experimental results that seem inconsistent with the model. The LSND experiment reported oscillations that cannot be accommodated within the 3$\nu$SM \cite{LSND} because the measured oscillation parameters do not agree with the ones measured by other experiments, differing by many orders of magnitude. The MiniBooNE experiment was build to address this anomalous result; however, MiniBooNE observed oscillation signals a low energy that cannot be understood within the 3$\nu$SM either \cite{MiniBooNE}.

The persistence of these results with the years has kept theorists busy trying to construct models to describe these anomalies. The possibility discussed here is that part or even all neutrino oscillation data could be described using the SME.

\section{Lorentz violation and neutrinos}

One simple reason to consider that Lorentz violation could describe neutrino oscillations is the interferometric nature of this phenomenon, which makes of these measurements sensitive probes of new physics. In fact, neutrino oscillations arise precisely near the Planck scale suppression expected for the coefficients in the SME. The measured values of the mass-square differences are $\Delta m^2_\odot\simeq10^{-22}$ GeV$^2$ and $\Delta m^2_\text{atm}\simeq10^{-21}$ GeV$^2$, which shows the high level of sensitivity achieved by neutrino oscillation experiments.

In the SME, free neutrinos are characterized by an effective hamiltonian describing propagation and oscillation of three active left-handed neutrinos and three active right-handed antineutrinos. The hamiltonian is represented by a $6\times6$ matrix, whose block-diagonal components can be decomposed into its Lorentz-invariant $(h_\text{LI})$ and Lorentz-violating parts $(h_\text{LV})$. The block describing neutrinos is a $3\times3$ matrix that can be written in the form $(h_\text{eff})_{ab}=(h_\text{LI})_{ab}+(h_\text{LV})_{ab}$, where the flavor indices span $a,b=e,\mu,\tau$. Here we see that the 3$\nu$SM is contained in the SME as a particular form of the Lorentz-invariant term $(h_\text{LI})_{ab}$. The elements of this effective hamiltonian for neutrinos are given by \cite{KM1}

\begin{equation}
\label{h_LV}
(h_\text{eff})_{ab} 
= E\,\delta_{ab}
+\frac{m^2_{ab}}{2E} 
+(a_L)^\alpha_{ab} \,\hat p_\alpha
-(c_L)^{\alpha\beta}_{ab} E \,\hat p_\alpha \hat p_\beta
+\ldots,
\end{equation}
where the first two terms correspond to $(h_\text{LI})_{ab}$, Lorentz violation in the minimal SME (renormalizable) is controlled by the next two terms, and the ellipsis denotes nonremormalizable terms corresponding to an infinite series in powers of the neutrino energy $E$ and direction of propagation $\hat p$ \cite{KP1995,SME,NRterms1,NRterms2}. The coefficients $m^2_{ab}$, $(a_L)^\alpha_{ab}$, and $(c_L)^{\alpha\beta}_{ab}$ are $3\times3$ matrices in flavor space, and the four-vector $\hat p_\alpha=(1;-\hat p)$ depends on neutrino direction of propagation. The first term in Eq. (\ref{h_LV}) is usually disregarded because oscillations depend on eigenvalue differences, which makes them insensitive to terms proportional to the identity. Despite the fact that these terms are irrelevant for oscillations they might be important for studies of stability and causality of the underlying theory \cite{stability}.

The corresponding $3\times3$ effective hamiltonian describing right-handed antineutrinos is obtained by replacing $m^2_{ab}\rightarrow m^2_{\bar a\bar b}=m^{2*}_{ab}$, $(a_L)^\alpha_{ab}\rightarrow(a_R)^\alpha_{\bar a\bar b}=-(a_L)^{\alpha*}_{ab}$, and $(c_L)^{\alpha\beta}_{ab}\rightarrow(c_R)^{\alpha\beta}_{\bar a\bar b}=(c_L)^{\alpha\beta*}_{ab}$, where the flavor indices span $\bar a,\bar b=\bar e,\bar\mu,\bar\tau$.

A third $3\times3$ matrix corresponding to the off-diagonal part of the $6\times6$ effective hamiltonian mixes neutrinos and antineutrinos and the structure is similar to that in Eq. (\ref{h_LV}) with the difference that the first two terms are absent because there is no Lorentz-invariant part. The coefficients for Lorentz violation $(a_L)$ and $(c_L)$ are replaced by other coefficients denoted $\tilde{H}$ and $\tilde{g}$. These coefficients lead to neutrino-antineutrino oscillations, a feature absent in the 3$\nu$SM.

The dependence of the effective hamiltonian in neutrino direction of propagation is a signal of the breakdown of invariance under rotations; nonetheless, there are coefficients for Lorentz violation that introduce only isotropic effects. If only the time components of the coefficients $(a_L)$ and $(c_L)$ are nonzero, then the general effective hamiltonian (\ref{h_LV}) becomes \cite{KM1}

\begin{equation}
\label{h_LV_FC}
(h_\text{eff})_{ab} 
= \frac{m^2_{ab}}{2E} 
+(a_L)^T_{ab} 
-\frac{4}{3}(c_L)^{TT}_{ab} E 
+\ldots,
\end{equation}
where only renormalizable terms are shown. This isotropic limit of the SME or {\it fried chicken} hamiltonian has been considered to study neutrinos since the early days of the SME \cite{SME,ColemanGlashow}. Nonrenormalizable terms represented by the ellipsis also appear
in the full SME and lead to increasing powers of the energy \cite{NRterms1}. 

In the even simpler case when the coefficients $(c_L)^{TT}_{ab}$ vanish, we obtain a hamiltonian for massive neutrinos modified by a constant potential $(a_L)^T_{ab}$. Notice the similarity of this scenario with the 3$\nu$SM in the presence of matter, in fact the components in flavor space of the coefficient for CPT-odd Lorentz violation $(a_L)^T_{ab}$ can be interpreted as unconventional corrections to matter effects, producing the same effects as non-standard matter interactions \cite{NSI}. Being CPT-odd, when neutrinos are replaced by antineutrinos these terms reverse the sign of their real parts $(a_L)^T_{ab}\to-(a_L)^{T*}_{ab}$, just like matter effects. To date, only $(a_L)^T_{e\mu}$ has been studied by LSND \cite{LSND_LV} and MiniBooNE \cite{MiniBooNE_LV}.

\newpage
\subsection{The KM plot}

The parametrization used by the 3$\nu$SM depends on three angles and two mass-squared differences, where the mixing angles determine the oscillation amplitude and the mass-squared differences control the oscillation phase. Current values of these oscillation parameters show that two-flavor scenarios are enough to describe most of the experiments to a good approximation. For this reason, the experimental results for the measurement of oscillation parameters are reported using a $\Delta m^2$-$\sin^22\theta$ plane. Even though this way to present results is useful because it shows how the wavelength of the oscillation (proportional to $\Delta m^2$) is related to the amplitude of the observed oscillation ($\sin^22\theta$), it assumes a particular energy dependence; in other words, the $\Delta m^2$-$\sin^22\theta$ plot is appropriate for a model in which the effective hamiltonian exhibits $1/E$ dependence only. For a more general description of neutrino propagation and oscillation that can incorporate Lorentz and CPT violation within a realistic effective field theory, unconventional energy dependence appears and the $\Delta m^2$-$\sin^22\theta$ plot becomes inappropriate. 

\begin{figure}[h!]
\centering
\includegraphics[width=70mm,angle=-90]{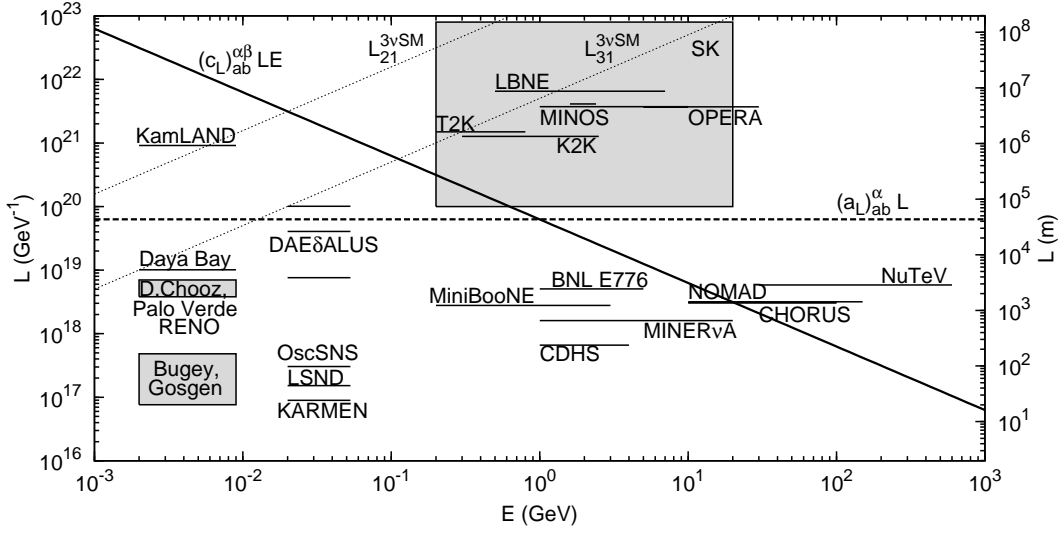}
\caption{The KM plot presenting the oscillation lengths that appear when Lorentz violation is included in the effective hamiltonian: the lines represent the points when the combinations $(c_L)^{\alpha\beta}_{ab}LE$ (solid line) and $(a_L)^\alpha_{ab}L$ (dashed line) become $2\pi$. The two parallel lines that grow with energy correspond to $L_{21}^\text{3$\nu$SM}$ and $L_{31}^\text{3$\nu$SM}$ (dotted lines).
} \label{Figure1_KM}
\end{figure}

In 2003, Kosteleck\'y and Mewes (KM) \cite{KM1} introduced a powerful model-independent plot that allows simultaneous
visualization of oscillation signals in different experiments. This KM plot applies to general models of neutrino oscillation
including ones with Lorentz and CPT violation. Figure~\ref{FigureKM} shows this KM plot, where the energy coverage $E$ and baseline $L$ of several experiments is presented. Notice that Eq. (\ref{P_ab}) can be written in the form

\begin{equation}\label{Eq1_OscProb}
P_{\nu_b\to\nu_a}=\delta_{ab}-4\sum_{a'>b'}U_{a'a}U_{a'b}U_{b'a}U_{b'b} \, \sin^2{\left(\frac{\pi L}{L_{a'b'}(E)}\right)},
\end{equation}
where we have written the oscillation amplitude in terms of the elements of the mixing matrix  $U_{a'a}$ and defined the oscillation length $L_{a'b'}(E)\equiv2\pi/|\Delta_{a'b'}(E)|$. For three neutrino flavors only two independent curves appear. 
Notice that the ratio $L/L_{a'b'}$ in Eq. (\ref{Eq1_OscProb}) controls the oscillation signal, in fact $L_{a'b'}/2$ physically corresponds to the location of the first maximum of a given oscillation channel. In other words, $L_{a'b'}/2$ indicates the minimal distance from the source to optimally measure neutrino appearance or disappearance. If the oscillation amplitude is appreciable, experiments that lie above the curve $L_{a'b'}/2$ will observe oscillations because the oscillation phase approaches to unity. Similarly, oscillation signals are suppressed for experiments that lie below the relevant curve. This is how meaningful information about oscillations can be visualized in the KM plot. For a given oscillation channel, explicit calculation of the oscillation probability is required to identify which of the two curves is relevant. In the 3$\nu$SM, the eigenvalue differences are $\Delta_{21}^\text{3$\nu$SM}=m^2_\odot/2E$ and $\Delta_{31}^\text{3$\nu$SM}=m^2_\text{atm}/2E$, which lead to the two straight lines that grow with the energy in Fig. \ref{Figure1_KM}. We can see that the disappearance of reactor antineutrinos in KamLAND is controlled by $L_{21}^\text{3$\nu$SM}$, whereas the disappearance of atmospheric and long-baseline neutrinos is controlled by $L_{31}^\text{3$\nu$SM}$. Notice that the KM plot is insensitive to sign of $\Delta_{a'b'}$, which in the 3$\nu$SM leads to our lack of knowledge about the neutrino mass hierarchy. In this plot we can also understand the problem of LSND and MiniBooNE oscillation signals for the 3$\nu$SM: these experiments are too far below the two mass lines; therefore, no signal of oscillation is expected in this model. One of the popular proposals to try to explain LSND and MiniBooNE results is the incorporation of sterile neutrino flavors, which would lead to new independent eigenvalue differences and therefore new lines in the KM plot. This possibility lies beyond the scope of this work.

In the case of Lorentz violation, the unconventional energy dependence introduced leads to more general eigenvalue differences and consequently to general curves in the KM plot. In the same way that the mass term generates the dimensionless combination $\Delta m^2_{ab}L/E$
that leads to oscillation lengths that grow linearly with energy, Lorentz-violating terms lead to the dimensionless combinations $(a_L)^\alpha_{ab}L$ and $(c_L)^{\alpha\beta}_{ab}LE$, which in the KM plot produce the other two curves shown in Fig. \ref{Figure1_KM}.

The drawing of these straight lines also provides information about the sensitivity of a given experiment to different coefficients for Lorentz violation. For instance, let us suppose that the Daya Bay reactor experiment finds no signals of some particular Lorentz-violating effect. This means that the $(a_L)$ coefficient responsible for that effect cannot be larger than a certain amount such that the signal of this coefficient for Lorentz violation is absent. That means that the corresponding line in the KM plot is far above of the experiment (at least one order of magnitude), similar to the horizontal line in Fig. \ref{Figure1_KM}, corresponding to an $(a_L)$ coefficient of order $10^{-20}$ GeV. A similar analysis applies to the $(c_L)$ coefficients using straight lines of negative slope instead.\footnote{the slope of lines associated with $(c_L)$ coefficients in the KM plot is -1} Notice that experiments with similar baseline are equally sensitive to $(a_L)$ coefficients, whereas the sensitivity to the effects of $(c_L)$ coefficients grows with the energy of the experiment. This means that future reactor experiments (Daya Bay, Double Chooz, and RENO) have sensitivities to $(a_L)$ coefficients similar to those of MiniBooNE and NuTeV; however, of this list NuTeV has the highest sensitivity to $(c_L)$ coefficients because of the higher energy. It is important, however, to remark that all coefficients have flavor indices that depend on the oscillation channel of study, which means that searches of different experiments are usually complementary, covering different sectors of the coefficient space.

\subsection{Observable signals of Lorentz violation}

Key features of Lorentz violation in neutrinos can be classified in six particular classes of signals that differ considerably from the 3$\nu$SM and can be studied in different experiments \cite{KM1}.

\subsubsection{Class I: spectral anomalies}
In Eq. (\ref{h_LV}) we find that non-negative powers of the energy appear in the effective hamiltonian. This unconventional energy dependence leads to oscillation phases that are energy independent or that grow with neutrino energy. Moreover, if more than one power of the energy appears, the mixing matrix becomes energy dependent. When this happens, general curves will represent the corresponding physics in the KM plot. 

These terms with unconventional energy dependence also introduce novel effects such as neutrinos that propagate at speeds different from the speed of light. In fact, neglecting flavor information and taking the isotropic limit (\ref{h_LV_FC}) for simplicity, the group velocity for neutrinos takes the form \cite{KP1995,SME,NRterms1,NRterms2,stability,AC,KR}

\begin{equation}\label{v_g}
v_g=1-\frac{m^2}{2E^2}+\mathaccent'27 c + \sum_{n=1}^\infty k^{(n)} E^n,
\end{equation} 
where $\mathaccent'27 c=-4c^{TT}/3$ and the quantities $k^{(n)}$ are proportional to isotropic coefficients for Lorentz violation associated to operators of mass dimension $n+4$. Notice that the mass term can only decrease the speed of propagation, making neutrinos move at speeds lower than the speed of light. Lorentz violation, on the other hand, can make neutrinos move either faster or slower than light. Particulary, at high energies the mass term becomes negligible and high-energy neutrinos could become superluminal. The SME thereby realizes the possible tachyonic nature of neutrinos proposed in 1985 \cite{tachyonic_nu}. Details about particle group velocities in the presence of Lorentz violation and the corresponding propagation at speeds other than that of light can be found in Refs. \cite{SME,stability,AC,KR}. Particular cases of Eq. (\ref{v_g}) have been studied in recent years \cite{DK1,DK2,KM1,KM2,OthDispRel,OthTachyonicNu,BMW,tandem,BLMW}. In the presence of Lorentz violation, neutrinos follow geodesics in a Finsler spacetime \cite{Finsler}.

\subsubsection{Class II: $E$-$L$ conflicts}
This class of signals refers to experimental results that challenge the 3$\nu$SM-based interpretation of oscillations produced by two independent mass-squared differences only. In the KM plot, this class refers to the requirement that experiments far below the two straight lines representing the oscillation lengths due to masses should not observe oscillation signals. The LSND and MiniBooNE anomalies correspond to observed signals of this type.

\subsubsection{Class III: periodic variations}
If different experiments search for the effects of coefficients for Lorentz violation, then a common frame of reference must be used, so that different results can be compared in a physically meaningful way. The adopted frame for reporting results of SME coefficients is centered in the Sun, whose $\hat Z$ axis directed north and parallel to the axis of rotation of the Earth, the $\hat X$ axis points toward the vernal equinox, and the $\hat Y$ axis completes the system. This Sun-centered celestial-equatorial frame \cite{KM_SunCFrame} is used in all searches of Lorentz violation based on the SME \cite{tables}.

Periodic variation of the oscillation probability is one of the key signals of Lorentz violation that arises when a nonzero coefficient for Lorentz violation has space components $\hat X$ or $\hat Y$, in which case for Earth-based experiments a sidereal modulation in the oscillation probability appears. Source and detector rotate with sidereal frequency $\omega_\oplus\simeq2\pi/(23\text{ h } 56\text{ min})$, making neutrino direction of propagation change with respect to the fixed nonzero coefficient. This time dependence appears in the oscillation probability as harmonics of the sidereal angle $\omega_\oplus T_\oplus$, which in the minimal SME takes the form

\begin{eqnarray}\label{P(OmegaT)}
P_{\nu_b\to\nu_a} &=&
(P_\mathcal{C})_{ab}+
(P_{\mathcal{A}_s})_{ab}\sin\omega_\oplus T_\oplus+
(P_{\mathcal{A}_c})_{ab}\cos\omega_\oplus T_\oplus\nonumber\\&&+
(P_{\mathcal{B}_s})_{ab}\sin2\omega_\oplus T_\oplus+
(P_{\mathcal{B}_c})_{ab}\cos2\omega_\oplus T_\oplus,
\end{eqnarray}
where the harmonic amplitudes depend on the different SME coefficients. Depending on the characteristics of the experiment and the oscillation channel of study up  to fourth harmonics can appear \cite{DKM2009,KM_SB}; nonetheless, higher harmonics are expected when nonrenormalizable terms are included \cite{NRterms1,NRterms2}. By measuring the harmonic amplitudes in Eq. (\ref{P(OmegaT)}) the effects of different coefficients can be tested. Annual variations can also appear; however, they are expected to be suppressed because of the nonrelativistic motion of the Earth around the Sun.

\subsubsection{Class IV: compass asymmetries}
This class of signals arise when a nonzero coefficient for Lorentz violation has space components in the $\hat Z$ direction of the Sun-centered frame. Coefficients with component in this direction produce no time dependence; nevertheless, since invariance under rotations is broken neutrino oscillations depend on the direction of propagation. 

\subsubsection{Class V: neutrino-antineutrino oscillations}
The SME describes neutrino-neutrino and antineutrino-antineutrino mixing through the coefficients $(a_L)$ and $(c_L)$; however, two other type of coefficients denoted by $\tilde{g}$ and $\tilde{H}$ mix neutrinos and antineutrinos, which allow the possibility of $\nu\leftrightarrow\bar\nu$ oscillations. The off-diagonal block of the $6\times6$ effective hamiltonian that depends on these coefficients has the form \cite{KM1}

\begin{equation}\label{h(g,H)}
h_{a\bar b} =
i\sqrt2 (\epsilon_+)_\alpha \tilde{H}^\alpha_{a\bar b} - i\sqrt2 (\epsilon_+)_\alpha\hat p_\beta \,\tilde g^{\alpha\beta}_{a\bar b} E,
\end{equation}
where $(\epsilon_+)_\alpha$ is a complex 4-vector representing the helicity state, and the flavor indices span $a=e,\mu,\tau$; $\bar b=\bar e,\bar\mu,\bar\tau$. These coefficients also introduce unconventional energy dependence and always cause direction-dependent effects, so the classes of signals described above can appear simultaneously. These coefficients can cause both lepton-number violating and lepton-number preserving oscillations, although their effects appear at second order. Details about the effects of these coefficients and their application to different experiments can be found in Ref. \cite{DKM2009}.

\subsubsection{Class VI: direct CPT violation}
Invariance of oscillations under CPT requires $P_{\nu_b\to\nu_a}=P_{\bar\nu_a\to\bar\nu_b}$ and if this condition is observed not to be satisfied is a clear indication of CPT violation and its corresponding Lorentz violation \cite{owg}. Notice, however, that this relation can be satisfied even if CPT violation exists.

\newpage
\subsection{Searches and models}

\subsubsection{Experimental searches}
For experiments in which large oscillation signals are expected according to the 3$\nu$SM, Lorentz violation can be incorporated as a perturbative effect. Details about the formalism and its application to different experiments can found in Ref. \cite{DKM2009}. In 2010, the MINOS experiment used this formalism to constrain a set of nine coefficients \cite{MINOS_LV2010}.

For experiments in which no oscillation signals are expected according to the 3$\nu$SM, Lorentz violation acts as the only cause of oscillations; not because neutrinos are massless but because the ratio $L/E$ is too small to generate a significant oscillation signal. The corresponding formalism is developed in Ref. \cite{KM_SB} and it can also be derived from the perturbative formalism in the limit of negligible neutrino masses \cite{DKM2009}. This formalism that can be applied to most of the short-baseline accelerator experiments was used by the LSND experiment in 2005 to perform the first search of sidereal variations in neutrino oscillations \cite{LSND_LV}. Later, in 2008, the MINOS experiment used its near detector as a short-baseline experiment to study the effects of another set of SME coefficients \cite{MINOS_LV2008}. In 2010, the IceCube experiment used very-high-energy neutrinos to constrain a subset of the coefficients studied in MINOS \cite{IceCube_LV}. IceCube used atmospheric neutrinos at an energy regime that neutrino masses are negligible, which makes this negligible-mass limit appropriate. More recently, the MiniBooNE experiment used the same technique to constrain combinations of SME coefficients \cite{MiniBooNE_LV}.

\subsubsection{Global models of neutrino oscillations}

After the development of the SME one of the natural questions was the possibility to construct models of neutrinos oscillations based on the SME consistent with all experimental data. Examples of these so-called global models are the bicycle model (2004) \cite{KM2}, the generalized bicycle model (2007) \cite{BMW}, the tandem model (2006) \cite{tandem}, the puma model (2010) \cite{DK1,DK2}, and the general isotropic bicycle model (2011) \cite{BLMW}. Other particular models can be found in Ref. \cite{more_models}. As the main focus of these proceedings is the puma model, we dedicate the next section to study this model in detail.

\section{The puma model}
Global descriptions of neutrino data usually take the 3$\nu$SM as starting point, which is then extended to incorporate new features by adding new parameters, forces, or particles. Based in the SME, the `puma' model \cite{DK1} is presented as an alternative to the 3$\nu$SM rather than an extension. This new model is characterized by an effective hamiltonian describing the oscillation of three active left-handed neutrinos in the form of a $3\times3$ matrix that in flavor space is given by

\begin{equation}
h_\text{eff}
= A\left(\begin{array}{ccc}
1 & 1 & 1 \\
1 & 1 & 1 \\
1 & 1 & 1 \end{array}\right) 
+ B\left(\begin{array}{ccc}
1 & 1 & 1 \\
1 & 0 & 0 \\
1 & 0 & 0 \end{array}\right) 
+ C\left(\begin{array}{ccc}
1 & 0 & 0 \\
0 & 0 & 0 \\
0 & 0 & 0 \end{array}\right),
\label{h_puma}
\end{equation}
where $A=m^2/2E$, $B=\mathaccent'27 aE^2$, and $C=\mathaccent'27 cE^5$ are functions of the neutrino energy $E$ and the quantities $\{m^2,\mathaccent'27 a,\mathaccent'27 c\}$ are the only three parameters of the model. Other models with the same texture as Eq. (\ref{h_puma}) but different powers of the energy can also be implemented that are globally consistent with established data \cite{DK2}.

This new model includes a mass term which controls oscillations at low energies in a very natural and symmetric form. Indeed, the effective hamiltonian at low energies is controlled by $A$ in (\ref{h_puma}), that in flavor space takes a democratic form. This low-energy term is contained in the Lorentz-invariant part of the neutrino sector in the SME, for this reason Lorentz and CPT invariance hold. Moreover, in flavor space this low-energy limit of the effective hamiltonian exhibits invariance under the action of the discrete group $S_3$, whose elements act on the three flavors $e,\mu,\tau$; in other words, there is no distinction between flavors in this limit. Even though there is no physical motivation for this indistinguishability in the first place, the many symmetries of the democratic form of $A$ make this assumption a well suited starting point to construct the effective hamiltonian. Exact diagonalization of the hamiltonian shows that in this limit the mixing becomes tribimaximal, this is a popular matrix in the neutrino literature proposed by Harrison, Perkins, and Scott \cite{TriBi} to describe neutrino mixing. Notice that the mixing matrix is completely given by the democratic form of the low-energy limit of the effective hamiltonian; in other words, no mixing angles are needed as degrees of freedom. In fact, at low energies the hamiltonian is characterized only by the mass parameter $m^2$. The tribimaximal mixing and the energy dependence at low energies guarantee the agreement of the model with solar and long-baseline reactor data.

\begin{figure}[h!]
\centering
\includegraphics[width=69mm,angle=-90]{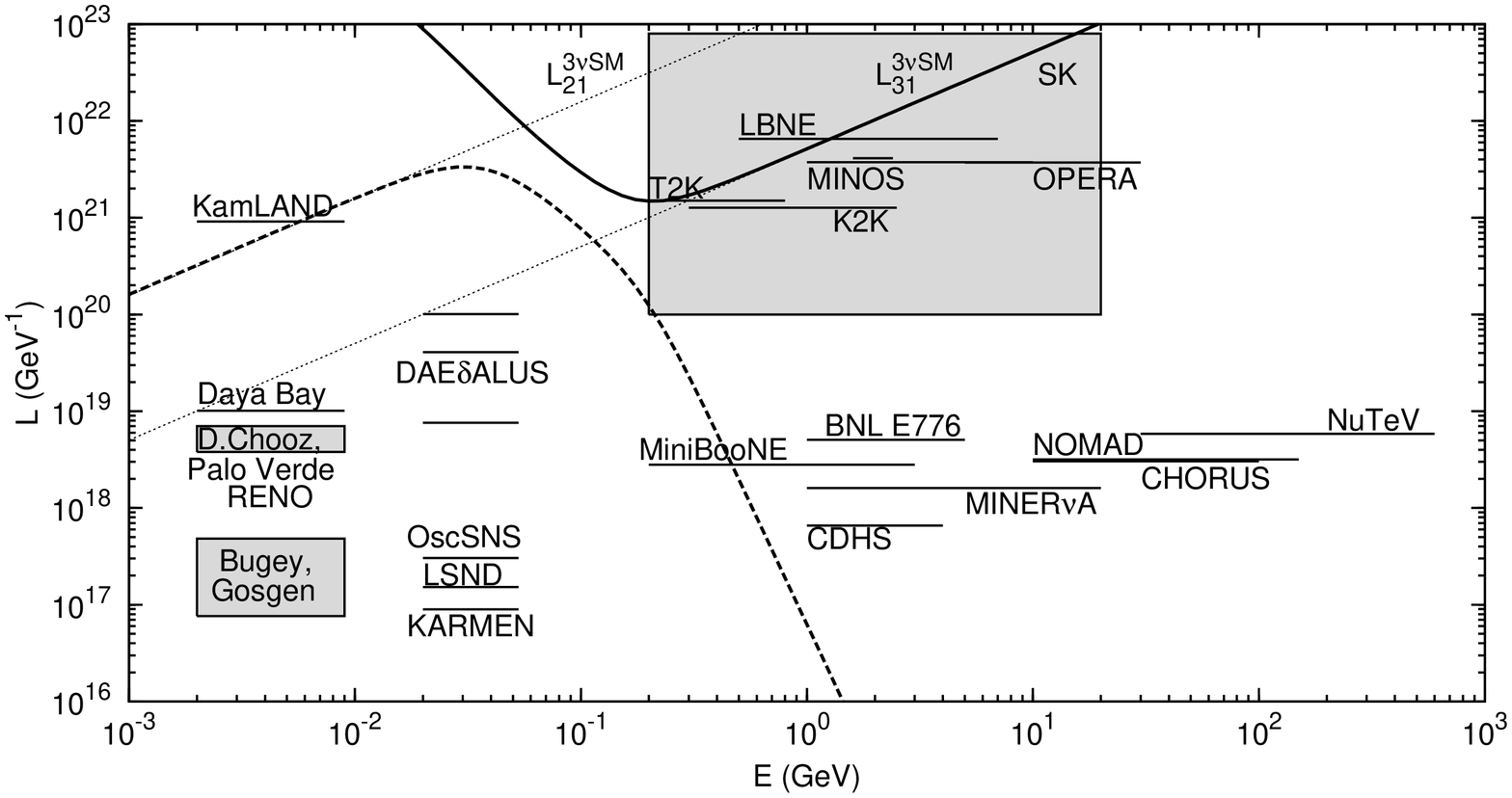}
\includegraphics[width=39mm,angle=-90]{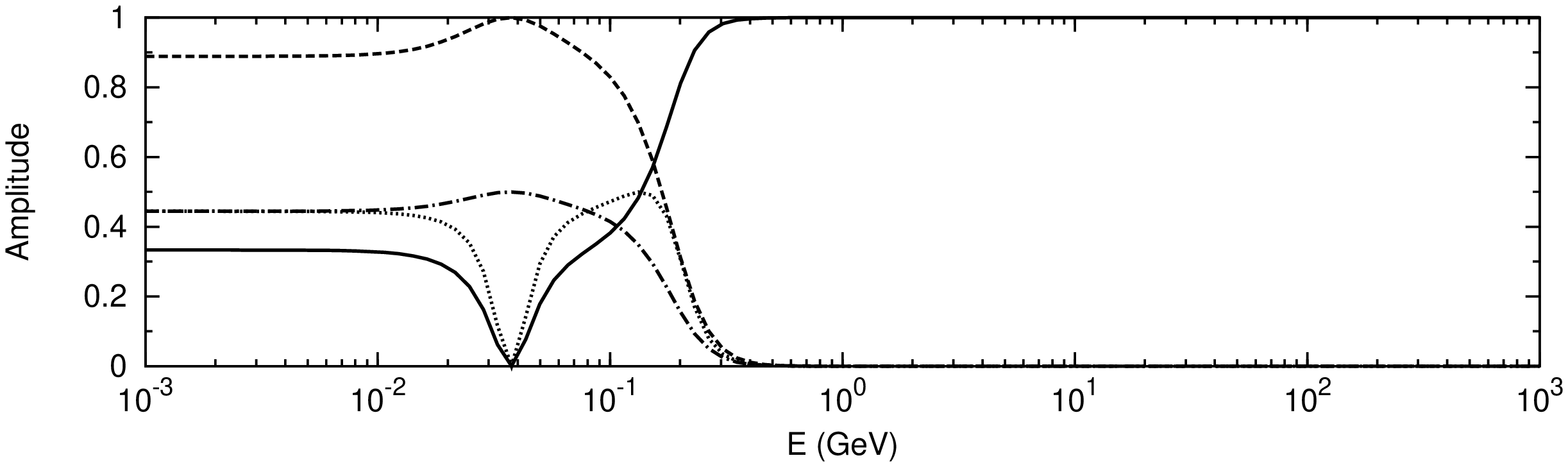}
\caption{Top: The KM plot indicating the behavior of the oscillation lengths $L_{31}$ (solid line) and $L_{21}$ (dashed line) as a function of neutrino energy. Straight lines represent the corresponding curves in the 3$\nu$SM (dotted line). Bottom: oscillation amplitudes for the oscillation channels $\nu_\mu\to\nu_\mu$ (solid line), $\bar\nu_e\to\bar\nu_e$ (dashed line), $\nu_\mu\to\nu_e$ (dotted line), and $\bar\nu_\mu\to\bar\nu_e$ (dash-dotted line).} \label{FigureKM}
\end{figure}

At high energies, the Lorentz-violating terms $B$ and $C$ dominate and the mass parameter becomes negligible. The different energy dependence of the two dominant terms leads to an energy-dependent mixing, which translates into oscillation amplitudes that vary with neutrino energy, a feature absent in the 3$\nu$SM. Notice that in Eq. (\ref{h_puma}) the terms $B$ and $C$ break not only Lorentz invariance but also the flavor symmetry $S_3$. This $S_3$ breaking is minimal, meaning that $S_3$ is broken to one of its $S_2$ subgroups. The preserved $S_2$ acts on the $\mu$-$\tau$ sector of the hamiltonian preserving the indistinguishability between these two flavors. This feature combined with the fast growing of the electron component of $h_\text{eff}$ produces maximal $\nu_\mu\leftrightarrow\nu_\tau$ mixing, in agreement with atmospheric and long-baseline accelerator measurements.
Despite the positive powers of the energy of the two terms at high energies in the effective hamiltonian, the fact that one entry in the hamiltonian grows faster than the other with neutrino energy triggers a Lorentz-violating seesaw mechanism \cite{KM1} that makes one of the eigenvalues behave like a mass because it becomes proportional to $1/E$. This eigenvalue controls the disappearance of atmospheric neutrinos and this energy dependence agrees with the oscillation signature observed by Super-Kamiokande as a function of $L/E$ \cite{SK(L/E)}. Another interesting feature of the puma model at high energies is the fact that the maximal $\nu_\mu\leftrightarrow\nu_\tau$ mixing mentioned above occurs when the seesaw mechanism is triggered. A consequence of this effect is that the mixing angle $\theta_{23}$ in the 3$\nu$SM is not necessary as a degree of freedom because the mixing is provided by the texture of the effective hamiltonian (\ref{h_puma}).

\subsection{KM plot for the puma model}

For the puma model, the KM plot is shown in Fig. \ref{FigureKM}. The plot provides a clear way to understand the main features of the model as well as its agreement with the description of established data at low and energies by the 3$\nu$SM. At low energy, the disappearance of reactor antineutrinos is controlled by $L_{21}$ that given the numerical value of the mass parameter coincides with the mass line in the 3$\nu$SM $L_{21}^\text{3$\nu$SM}$. Since the $L$-$E$ plane only gives information about the size of the oscillation phase, the oscillation amplitudes for the most important oscillation channels are plotted underneath. For reactor experiments, we see that for antineutrinos within the reactor spectrum (2-9 MeV) the amplitude remains constant and equal to 8/9 (from the tribimaximal mixing). The amplitude goes to maximal and then decreases; however, this occurs at energies beyond the reactor spectrum. Notice that the null signals in short-baseline reactor experiments such as CHOOZ, Palo Verde, Bugey, and G\"osgen in the puma model occur because they are too far below $L_{21}$ in the KM plot. In the contrary, in the 3$\nu$SM this is a consequence of a small oscillation amplitude determined by $\theta_{13}$.

Positive powers of the energy produce decreasing curves in the KM plot; nevertheless, the seesaw mechanism \cite{KM1,KM2} makes one of the curves mimic the effects of a mass by approaching to the straight line $L_{31}^\text{3$\nu$SM}$ for energies above 1 GeV. Since the curve $L_{31}$ controls the survival of $\nu_\mu$ at high energies, no difference appears between the puma and 3$\nu$SM at this energy scale. Moreover, the oscillation amplitude approaches to maximal as the two curves $L_{21}$ and $L_{31}$ separate, as presented in the amplitudes plot. This plot also shows that only the amplitude $\nu_\mu\to\nu_\tau$ is significant at energies above 1 GeV; this is consistent with the null results of several high-energy short-baseline accelerator experiments. The null results in CHORUS and NOMAD in the $\nu_\mu\to\nu_\tau$ channel is a consequence of being too far below $L_{31}$.

One of the remarkable features of the puma model occurs in the middle energy range, where the MiniBooNE experiment lies in the KM plot. Direct calculation of the $\nu_\mu\to\nu_e$ oscillation channel shows that this oscillation is controlled by $L_{21}$. In the KM plot we see that when the seesaw mechanism makes $L_{31}$ grow linearly with energy and behaving like a mass, the other oscillation length $L_{21}$ abruptly decreases with energy. The KM plot shows how this curve runs over MiniBooNE, producing a large oscillation phase in this experiment. The oscillation amplitude, on the other hand, is large and decreases for energies above 500 MeV. The combined effect is a significant oscillation signal in MiniBooNE but only at low energies, which is precisely the effect observed in this experiment that cannot be accommodated by the 3$\nu$SM \cite{MiniBooNE}. Oscillations signals in MiniBooNE were first predicted by other models based in the SME \cite{tandem}.

\subsection{Predictions}

The unconventional energy dependence of the puma model produces oscillation amplitudes that vary with energy. The tribimaximal mixing at low energies takes a complicated form at high energies that can produce $\nu_\mu\to\nu_e$ signals, a feature absent in the 3$\nu$SM. Since the mixing is tribimaximal at low energies, no significant disappearance signals are expected in future short-baseline reactor experiments because they are too far below $L_{21}$ in the KM plot. For the same reason, only Daya Bay could observe oscillation signals driven by the solar mass-square difference. In the puma model, long-baseline experiments studying $\nu_\mu\to\nu_e$ oscillations are similar to MiniBooNE; therefore, significant appearance signals in this channel are expected a low energies. In other words, the size of the oscillation signal\footnote{determined by $\sin^22\theta_{13}$ in the 3$\nu$SM.} should be larger in T2K compared to MINOS or NO$\nu$A due to the different neutrino energy. Recent results from MINOS and T2K are compatible with this prediction \cite{T2K_MINOS}.

Even though the model contains one mass parameter, no predictions can be made about the absolute neutrino masses because in the effective hamiltonian (\ref{h_puma}) terms proportional to identity have been disregarded.

An important anomalous result is the LSND signal \cite{LSND} that has not been addressed here. In Ref. \cite{DK2} details are described for a simple non-polynomial extension of the model presented here that includes LSND.

\bigskip 

\begin{thebibliography}{9}   


\bibitem{pdg}
K.\ Nakamura {\it et al.}, J.\ Phys.\ G {\bf 37}, 075021 (2010).

\bibitem{NeutralMesons}
V.A.\ Kosteleck\'y and R.\ Van\ Kooten,
Phys.\ Rev.\ D {\bf 82}, 101702(R) (2010);
V.A.\ Kosteleck\'y,
Phys.\ Rev.\ Lett.\ {\bf 80}, 1818 (1998).

\bibitem{KS89}   
  V.A.\ Kosteleck\'y and S.\ Samuel,
  Phys.\ Rev.\  D {\bf 39}, 683 (1989);
  V.A.\ Kosteleck\'y and R.\ Potting,
  Nucl.\ Phys.\  B {\bf 359}, 545 (1991).
  
\bibitem{DK1}
 J.S.\ D\'\i az and V.A.\ Kosteleck\'y,
 Phys.\ Lett.\ B {\bf 700}, 25 (2011). 

\bibitem{DK2}
 J.S.\ D\'\i az and V.A.\ Kosteleck\'y,
 arXiv:1108.1799.
  
\bibitem{KP1995}
  V.A.\ Kosteleck\'y and R.\ Potting,
  Phys. Rev. D 51, 3923 (1995). 
   
\bibitem{SME}
D.\ Colladay and V.A.\ Kosteleck\'y, 
Phys.\ Rev.\ D {\bf 55}, 6760 (1997);
Phys.\ Rev.\ D {\bf 58}, 116002 (1998).

\bibitem{SME2004} 
V.A.\ Kosteleck\'y,
Phys.\ Rev.\ D {\bf 69}, 105009 (2004).

\bibitem{NRterms1}
V.A.\ Kosteleck\'y and M.\ Mewes, 
in preparation.

\bibitem{NGmodes}
  V.A.\ Kosteleck\'y and R.\ Potting,
  Gen.\ Rel.\ Grav.\  {\bf 37}, 1675 (2005);
  %
  Phys.\ Rev.\  {\bf D79}, 065018 (2009);
  %
  S.M.\ Carroll {\it et al.},
  Phys.\ Rev.\ D {\bf 80}, 025020 (2009); 
  %
  N. Arkani-Hamed {\it et al.},
  JHEP {\bf 0507}, 029 (2005);
  %
  O.\ Bertolami and J.\ Paramos,
  Phys.\ Rev.\  D {\bf 72}, 044001 (2005);
  %
  R.\ Bluhm and V.A.\ Kosteleck\'y,
  Phys.\ Rev.\  D {\bf 71}, 065008 (2005);
  %
  R.\ Bluhm {\it et al.},
  Phys.\ Rev.\ D {\bf 77}, 065020 (2008). 
  %
  V.A.\ Kosteleck\'y and S.\ Samuel,
  Phys.\ Rev.\  D {\bf 40}, 1886 (1989);
  %
  Phys.\ Rev.\ Lett.\  {\bf 63}, 224 (1989);
  %
  M.D.\ Seifert,
  Phys.\ Rev.\  D {\bf 79}, 124012 (2009).

\bibitem{owg}
O.W.\ Greenberg,
Phys.\ Rev.\ Lett.\  {\bf 89}, 231602 (2002);
arXiv:1105.0927.

\bibitem{tables}
{\it Data Tables for Lorentz and CPT Violation,}
V.A.\ Kosteleck\'y and N.\ Russell,
Rev.\ Mod.\ Phys. {\bf 83}, 11 (2011).

\bibitem{SME_developments}
  B.\ Altschul {\it et al.}, 
  Phys.\ Rev.\ D\ {\bf 81}, 065028 (2010). 
  %
  V.A.\ Kosteleck\'y, R.\  Lehnert, and M.\ Perry,
  Phys. Rev. D 68, 123511 (2003);
  %
  Q.\ Bailey and V.A.\ Kosteleck\'y,
  Phys.\ Rev.\ D\ {\bf 74}, 045001 (2006);
  %
  V.A.\ Kosteleck\'y, N.\ Russell, and J.\ Tasson,
  Phys.\ Rev.\ Lett.\ {\bf 100}, 111102 (2008);
  %
  V.A.\ Kosteleck\'y and J.\ Tasson,
  Phys.\ Rev.\ Lett.\  {\bf 102}, 010402 (2009);
  %
  Phys.\ Rev.\ D\ {\bf 83}, 016013 (2011).
  
\bibitem{KamLAND(L/E)} 
KamLAND Collaboration, 
T.\ Araki {\it et al.}, 
Phys.\ Rev.\ Lett.\ {\bf 94}, 081801 (2005);
S.\ Abe {\it et al.}, 
Phys.\ Rev.\ Lett.\ {\bf 100}, 221803 (2008).

\bibitem{SK(L/E)}   
  Super-Kamiokande Collaboration, Y.\ Ashie {\it et al.}, 
  Phys.\ Rev.\ Lett.\  {\bf 93}, 101801 (2004).  
   
\bibitem{LSND}   
LSND Collaboration, A.\ Aguilar {\it et al.},
Phys.\ Rev.\ D {\bf 64}, 112007 (2001).

\bibitem{MiniBooNE}   
  MiniBooNE Collaboration, A.A.\ Aguilar-Arevalo, {\it et al.}, 
  Phys.\ Rev.\ Lett.\  {\bf 98}, 231801 (2007);
  Phys.\ Rev.\ Lett.\  {\bf 102}, 101802 (2009);
  Phys.\ Rev.\ Lett.\  {\bf 105}, 181801 (2010).
  
\bibitem{KM1}
V.A.\ Kosteleck\'y and M.\ Mewes, 
Phys.\ Rev.\ D {\bf 69}, 016005 (2004).

\bibitem{NRterms2}
V.A.\ Kosteleck\'y and M.\ Mewes, 
Phys.\ Rev.\ D {\bf 80}, 015020 (2009);
%
Astrophys.\ J.\ Lett.\ {\bf 689}, L1 (2008). 

\bibitem{stability}
V.A.\ Kosteleck\'y and R.\  Lehnert,
Phys. Rev. D {\bf 63}, 065008 (2001).   

\bibitem{ColemanGlashow}
S.\ Coleman and S.L.\ Glashow, 
Phys.\ Rev.\ D {\bf 59}, 116008 (1999). 

\bibitem{NSI}
Y.\ Grossman, 
Phys.\ Lett.\ B 359, 141 (1995). 

\bibitem{LSND_LV}
LSND Collaboration, L.B.\ Auerbach {\it et al.},
Phys.\ Rev.\ D {\bf 72}, 076004 (2005).

\bibitem{MiniBooNE_LV}
MiniBooNE Collaboration, A.A.\ Aguilar-Arevalo, {\it et al.}, 
arXiv:1109.3480. 

\bibitem{AC}
B.\ Altschul and D.\ Colladay, 
Phys.\ Rev.\ D {\bf 71}, 125015 (2005).

\bibitem{KR}
  V.A.\ Kosteleck\'y and N.\ Russell,
  Phys. Lett. B 693, 443 (2010). 
  
\bibitem{tachyonic_nu}
A.\ Chodos {\it et al.},
Phys.\ Lett.\ B 150, 431 (1985);
%
Mod.\ Phys.\ Lett.\ A7, 467 (1992);
%
A.\ Chodos and V.A.\ Kosteleck\'y, 
Phys.\ Lett.\ B 336, 295 (1994);
%
V.A.\ Kosteleck\'y,
in {\it Topics on Quantum Gravity and Beyond}, edited by F.\ Mansouri and J.J.\ Scanio (World Scientific, Singapore, 1993). 
  
\bibitem{KM2}
V.A.\ Kosteleck\'y and M.\ Mewes, 
Phys.\ Rev.\ D {\bf 70}, 031902 (R) (2004).

\bibitem{BMW}
  V.\ Barger, D.\ Marfatia and K.\ Whisnant,
  Phys.\ Lett.\  B {\bf 653}, 267 (2007).

\bibitem{tandem}
T.\ Katori {\it et al.}, 
Phys.\ Rev.\ D {\bf 74}, 105009 (2006).

\bibitem{BLMW}
  V.\ Barger, J.\ Liao, D.\ Marfatia and K.\ Whisnant,
  Phys.\ Rev.\  D, in press, arXiv:1106.6023.

\bibitem{OthTachyonicNu}
M.J.\ Radzikowski,
in {\it CPT and Lorentz Symmetry V}, edited by V.A.\ Kosteleck\'y (World Scientific, Singapore, 2011);
%
N.-P.\ Chang,    
hep-ph/0410175;
%
R.\ Ehrlich,
Phys.\ Rev.\ D {\bf 60} 017302, (1999);
%
S.\ Giani,
CERN-OPEN-97-036, hep-ph/9712265. 

\bibitem{OthDispRel}
A.\ Sakharov, J.\ Ellis, N.\ Harries, A.\ Meregaglia, and A.\ Rubbia,
J.\ Phys.\ Conf.\ Ser.\ 171, 012039 (2009);
%
D.M.\ Mattingly, L.\ Maccione, M.\ Galaverni, S.\ Liberati, and G.\ Sigl, 
JCAP 1002, 007 (2010). 

\bibitem{Finsler}
  V.A.\ Kosteleck\'y,
  Phys. Lett. B 701, 137 (2011).

\bibitem{KM_SunCFrame}
  V.A.\ Kosteleck\'y and M.\ Mewes,
  Phys.\ Rev.\  D {\bf 66}, 056005 (2002).

\bibitem{DKM2009}
J.S.\ D\'iaz {\it et al.},
Phys.\ Rev.\ D {\bf 80}, 076007 (2009).

\bibitem{KM_SB}
V.A.\ Kosteleck\'y and M.\ Mewes,
Phys.\ Rev.\ D {\bf 70}, 076002 (2004).

\bibitem{MINOS_LV2010}
MINOS Collaboration, P.\ Adamson {\it et al.},
Phys.\ Rev.\ Lett.\ {\bf 105}, 151601 (2010).

\bibitem{MINOS_LV2008}
MINOS Collaboration, P.\ Adamson {\it et al.},
Phys.\ Rev.\ Lett.\ {\bf 101}, 151601 (2008).

\bibitem{IceCube_LV}
IceCube Collaboration, R.\ Abbasi {\it et al.},
Phys.\ Rev.\ D {\bf 82}, 112003 (2010).
  
\bibitem{more_models}
N.\ Cipriano Ribeiro {\it et al.},
Phys.\ Rev.\ D {\bf 77}, 073007 (2008);
A.E.\ Bernardini and O.\ Bertolami,
Phys.\ Rev.\ D {\bf 77}, 085032 (2008);
B.\ Altschul,
J.\ Phys.\ Conf.\ Ser. {\bf 173} 012003 (2009);
S.\ Hollenberg, O.\ Micu, and H.\ P\"as,
Phys.\ Rev.\ D {\bf 80}, 053010 (2009);
S.\ Ando, M.\ Kamionkowski, and I.\ Mocioiu,
Phys.\ Rev.\ D {\bf 80}, 123522 (2009);
M.\ Bustamante {\it et al.}, 
J.\ Phys.\ Conf.\ Ser.\ {\bf 171}, 012048 (2009);
S.\ Yang and B.-Q.\ Ma,
Int.\ J.\ Mod.\ Phys.\ A {\bf 24}, 5861 (2009);
P.\ Arias and J.\ Gamboa,
Int.\ J.\ Mod.\ Phys.\ A {\bf 25}, 277 (2010);
A.\ Bhattacharya {\it et al.},
JCAP {\bf 1009}, 009 (2010);
C.\ Liu, J.-t.\ Tian, and Z.-h.\ Zhao,
Phys.\ Lett.\ B {\bf 702}, 154 (2011).
  
\bibitem{TriBi} 
  P.F.\ Harrison, D.H.\ Perkins, and W.G.\ Scott,
  Phys. Lett. B {\bf 530}, 167 (2002).
  
\bibitem{T2K_MINOS}
T2K Collaboration,
K.\ Abe {\it et al.},
Phys.\ Rev.\ Lett.\ {\bf 107}, 041801 (2011);
MINOS Collaboration,
P.\ Adamson {\it et al.},
arXiv:1108.0015.






\end{thebibliography}

\end{document}